\documentclass{article}

\usepackage{microtype}
\usepackage{graphicx}
\usepackage{subfigure}
\usepackage{booktabs} 
\usepackage{mathtools, nccmath}
\usepackage{amsmath}
\usepackage{amssymb}
\usepackage{colortbl}
\usepackage{stix}
\usepackage{fancyhdr}
\usepackage{pifont}
\usepackage[colorlinks,allcolors=black,pdftex]{hyperref}

\newcommand{\JM}[1]{\textcolor{black}{#1}}

\DeclarePairedDelimiter{\nint}\lfloor\rceil

\newcommand{\cmark}{\ding{51}}%
\newcommand{\xmark}{\ding{55}}%

\definecolor{ora}{rgb}{0.97,0.88,0.84}
\definecolor{commentcolor}{RGB}{110,154,155}   

\usepackage[accepted]{mlsys2024}

\mlsystitlerunning{Torch2Chip}

\begin{document}

\twocolumn[
\mlsystitle{Torch2Chip: An End-to-end Customizable Deep Neural Network Compression and Deployment Toolkit for Prototype Hardware Accelerator Design}




\begin{mlsysauthorlist}
\mlsysauthor{Jian Meng}{to}
\mlsysauthor{Yuan Liao}{to}
\mlsysauthor{Anupreetham Anupreetham}{to}
\mlsysauthor{Ahmed Hasssan}{to}
\mlsysauthor{Shixing Yu}{to}
\mlsysauthor{Han-sok Suh}{to}
\mlsysauthor{Xiaofeng Hu}{to}
\mlsysauthor{Jae-sun Seo}{to}
\end{mlsysauthorlist}

\mlsysaffiliation{to}{Department of Electrical and Computer Engineering, Cornell University, USA}
\mlsyscorrespondingauthor{Jae-sun Seo}{js3528@cornell.edu}

\mlsyskeywords{Machine Learning, MLSys}

\vskip 0.3in

\begin{abstract}
Deep neural network~(DNN) compression~(e.g., quantization, pruning) has been widely investigated in various deep learning tasks~(e.g., vision and language). 
The development of model compression is continuously motivated by the evolution of various neural network accelerator designs with ASIC or FPGA. On the algorithm side, the ultimate goal of quantization or pruning is accelerating the expensive DNN computations on low-power hardware. 
However, such a ``design-and-deploy'' workflow faces under-explored challenges in the current hardware-algorithm co-design community due to some unavoidable flaws.
First, 
although the state-of-the-art quantization algorithm can achieve ultra-low precision with negligible degradation of accuracy, the latest deep learning framework~(e.g., PyTorch) can only support non-customizable 8-bit precision, data format, and parameter extraction workflow for CNN. Secondly, the ultimate goal of quantization is to enable the computation with low-precision data~(e.g., 4-bit integer). However, the current SoTA algorithm treats the quantized integer as an intermediate result, while the final output of the quantizer is the ``discretized'' floating-point values, ignoring the practical needs and adding additional workload to hardware designers for integer parameter extraction and layer fusion.
Finally, the compression toolkits designed by the industry are constrained to their in-house product or a handful of algorithms. The limited degree of freedom in the current toolkit and the under-explored customization hinder the prototype ASIC or FPGA-based accelerator design. To resolve these challenges, we propose Torch2Chip, an open-sourced, fully customizable, and high-performance toolkit that supports the user-designed compression algorithm followed by automatic model fusion and parameter extraction. Torch2Chip incorporates the hierarchical design workflow, and the user-customized compression algorithm will be directly packed into the deployment-ready format for either prototype chip verification with either CNN or vision transformer~(ViT). Furthermore, Torch2Chip covers a wide range of training methods to achieve high performance, from basic supervised learning to state-of-the-art (SoTA) lightweight self-supervised learning~(SSL). 
Code is available at \url{https://github.com/SeoLabCornell/torch2chip}.

\end{abstract}
]



\printAffiliationsAndNotice{} 

\section{Introduction}
\label{sec:intro}
Deep neural network compression has been developed as a critical recipe and almost mandatory step for energy-efficient deep learning. Starting from the early exploration of magnitude-based pruning~\citep{han2015deep} and low-precision quantization~\citep{zhou2016dorefa}, a variety of pruning/quantization algorithms have been presented 
along the rapid evolution in DNN architectures, across CNNs~\citep{liu2021sparse, yang2020harmonious}, transformers~\citep{yu2022unified, li2023vit}, self-supervised learning~\citep{meng2023slimmed} models, etc., generating compact models while preserving high accuracy.

To largely reduce hardware storage, computation and energy, 
hardware designers are motivated to adopt the state-of-the-art (SoTA) pruning and quantization schemes for custom DNN hardware accelerator designs, including digital ASIC~\citep{Lee_JSSC19,Moon_ISSCC23,Conti_ISSCC23,Desoli_ISSCC23}, FPGA~\citep{Yang_FPGA19, meng2021fixyfpga}, or analog compute-in-memory platforms~\cite{pimca, Huang_ISSCC23}. Besides the customized accelerator design, industry-standard compression toolkits~\citep{gorbachev2019openvino, siddegowda2022neural} are developed to support the off-the-shelf commercial hardware platforms for automated compression and deployment.

\begin{figure*}[t!]
    \centering
    \includegraphics[width=0.9\linewidth]{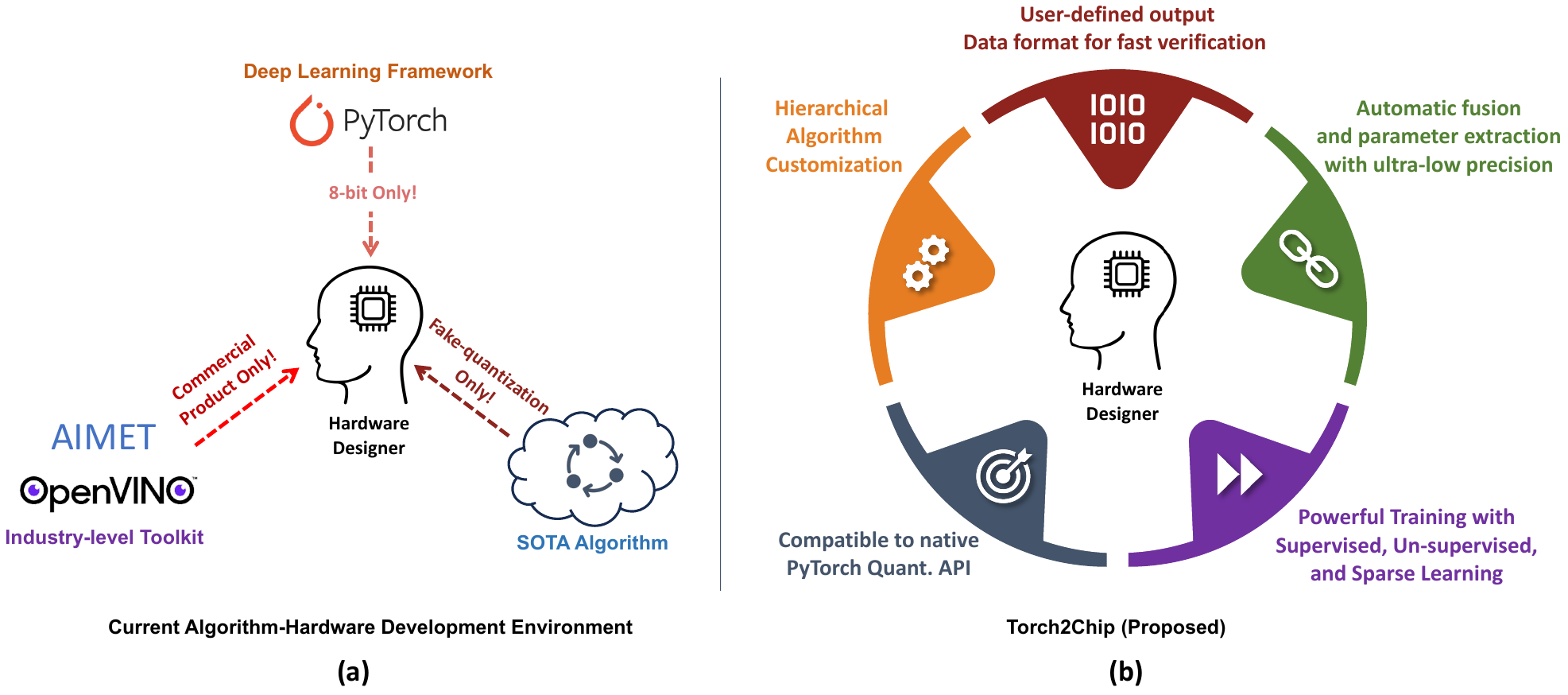}
    \caption{(a) Current algorithm-hardware development environment with disconnected workflow. (b) Proposed systematic toolkit \texttt{Torch2Chip} with high degree of customization versatility.}
    \label{fig:fig1}
\end{figure*}

Although significant improvements have been achieved in DNN compression algorithms and energy-efficient hardware accelerators, the hardware designer faces unavoidable challenges in ``compress-and-deploy'' workflow due to the following reasons:

\paragraph{Gap between ML Framework and Hardware Designer.}
Motivated by the necessity of model compression, the popular deep learning~(DL) framework~(e.g., PyTorch~\citep{NEURIPS2019_9015}) has released the dedicated compression API. However, the built-in quantization workflow is not the best candidate for hardware designers. In particular, the data precision is fixated as 8-bit with almost zero degree of customization. Although the compressed model is suitable for CPU or GPU-based acceleration, the pre-fixed, 8-bit only data format lags behind the ultra-low precision quantization algorithm and prototype hardware accelerators.  The user-customized training method or post-training calibration scheme cannot be implemented directly inside the API for sub-8-bit precision. Meanwhile, prototype hardware accelerators use various precision and algorithm customizations~\cite{Lee_JSSC19, Moon_ISSCC23}. 
The non-customizable compression scheme of the DL framework is disconnected with the needs of hardware designers.

\paragraph{Gap between SoTA Algorithm and Hardware Designer.}
On the algorithm side, most of the recent quantization algorithms merge the ``quantize-dequantize'' together to ensure the correct gradient propagation~\citep{li2021brecq, xiao2023smoothquant, wei2022qdrop}. As a result, the quantized low-precision integer is not the final output of the quantizer, and it cannot be accessed directly with the pre-trained model. As a result, fusing the normalization layer~\citep{jacob2018quantization} and extracting the quantized parameters requires additional algorithm manipulation from the hardware designer. \JM{Recent efforts on vision model quantization focus on efficient post-training quantization~(PTQ) with noise-robust biasing~\citep{liu2023noisyquant} and efficient dequantization~\cite{han2023abcd}. 
However, the compressed model is saved in full floating-point precision in the PyTorch model file, and the compatibility of further hardware deployment is ignored by default. }
Therefore, the SoTA quantization algorithms and hardware designers are disconnected due to the absence of the systematic ``customize-and-deploy'' workflow.  

\paragraph{Gap between Industry-level Toolkit and Prototyping.}
In addition to the native compression workflow of the DL frameworks, industry-standard hardware-algorithm co-design toolkits~\citep{gorbachev2019openvino, siddegowda2022neural} have been recently proposed for compression with the customized algorithm. 
However, almost all of the industry-level toolkits are designed for the dedicated commercial product~(e.g., CPU, \JM{Neural Compute Stick}~\citep{xu2017convolutional}) rather than customized chips. Specifically, the low-level RTL verification requires properly unrolled weight tensors with hexadecimal or binary data format in the memory module. With the fixed data precision~\cite{gorbachev2019openvino} and limited quantization options~\citep{siddegowda2022neural}, the industry-standard toolkits are sub-optimal for custom hardware that aim to adopt SoTA quantization algorithms.

From the perspectives of the hardware designers, the conflicts from the DL framework, SoTA algorithm, and current toolkits formulate the cumbersome designation workflow of chip prototyping, as shown in Figure~\ref{fig:fig1}(a). Motivated by that, the following question arises:

\textit{How to automate the deployment process from model compression to the prototype chip deployment, where the automatic workflow supports the fully customized SoTA compression method and training scheme?}

We attempt to answer the above question from the following perspectives that are embedded into the proposed \texttt{Torch2Chip} toolkit:

\paragraph{1) Hierarchical customized quantization build-up.}
Most open-sourced quantization algorithms~\citep{librecq,liunonuniform} perform ``quantization-dequantization'' all at once in a \textbf{single} computation path for both training or inference, \JM{and it is difficult to separate the integer-only computation from the custom quantizer.}
In this work, \texttt{Torch2Chip} separates the training and inference path with separate computation graphs, to perform the fake-quantized computation for training and quantized integer-only operation for inference. 
The proposed ``Dual-Path'' design is employed as the bottom-level logic for \texttt{Base quantizer} and \texttt{Base layers}, \JM{which avoids interference between training and inference while keeping the freedom of customization in training.} 
On top of that, the user-defined algorithms are constructed in a hierarchical fashion by computing the quantization parameter properly for either quantization-aware training~(QAT) or post-training quantization~(PTQ). In other words, \texttt{Torch2Chip} \textbf{only} requires users to design the training path and update the quantization parameter properly, while the remaining conversion will be completed automatically and executed in the inference path. 
\paragraph{2) Automatic fusion and parameter extraction for ultra-low precision.}
BatchNorm~(BN) fusion~\citep{jacob2018quantization} has been employed as a mandatory step for the post-training process in various toolkits. Normalization parameters are assumed to be fused into weights before quantization. 
However, prior works have shown the instability and degraded performance caused by the BN fusion with ultra-low-precision quantization~($\leq$8-bit)~\citep{park2020profit}. 
As a result, BN parameters have to be fused into channel-wise scaling and shifting. Unfortunately, the channel-wise fusion is \textbf{not} fully supported by PyTorch~\citep{NEURIPS2019_9015}, even though it has been employed in the low-precision accelerator design~\citep{meng2021fixyfpga, zhao2019automatic}. 
Motivated by that, \texttt{Torch2Chip} automates the fusion and parameter extraction in both weight-based BN fusion~(for 8-bit) and channel-wise scaling~(for sub-8-bit). 


\paragraph{3) Unsupervised foundation model training for powerful model compression and deployment.}
In addition to the highly versatile post-compression conversion and extraction, \texttt{Torch2Chip} also incorporates a wide range of training methods. Recent works on self-supervised learning~\cite{zbontar2021barlow, bardes2022vicregl, meng2023slimmed} have demonstrated stronger transfer learning performance than supervised learning. The learned superior visual representation becomes a powerful foundation for downstream vision tasks, which widely exists in resource-constrained edge computing. Different from the industry-level toolkit~\cite{gorbachev2019openvino, siddegowda2022neural} with supervised model trainer only, \texttt{Torch2Chip} provides the SoTA contrastive self-supervised learning method~\cite{meng2023slimmed} that is designed for lightweight~(e.g., MobileNet) encoders. In particular, the 8-bit, deployment-ready MobileNet-V1 can achieve 94.37\% and 74.29\% accuracy on CIFAR-10 and CIFAR-100 datasets, outperforming the conventional supervised training by a significant margin.  

\paragraph{4) User-defined output data format for fast deployment}
Different from the conventional integer data format in PyTorch~(\texttt{torch.qint8}) or Numpy~(\texttt{np.int8}), 
hardware description language (HDL) such as Verilog and SystemVerilog
only supports the raw Hexadecimal or Binary values. Ignoring the data format mismatch introduces additional conversion effort to the verification process of the prototyped chip. In this work, \texttt{Torch2Chip} exports the model with various formats, including raw integer tensors in PyTorch output file, hardware-readable Hexadecimal raw data, and the PyTorch native integer format~(\texttt{torch.qint8}). 

\section{Related Work}
\subsection{Quantization}\label{sec:rel_quant}
Quantization has been widely investigated together with the evolution of deep learning. Starting from the early ResNet~\cite{he2016deep} all the way to the recent large language models~\citep{narayanan2021efficient}, the elevation of the model complexity keeps necessitating efficient compression algorithms. To that end, various quantization methods~\citep{nagel2020up, li2021brecq, wei2022qdrop, xiao2023smoothquant} have been proposed together with the dedicated prototype accelerator~\cite{yin2020xnor, he20202, wang2023digital} among different hardware platforms. The core computation of quantization can be generalized as a) Data format scaling, b) Low precision computation, and c) Output rescaling. Mathematically, given the weights $\mathcal{W}^l$ and input activation $\mathcal{X}^l$ of a layer $l$, the quantization process can be expressed as:
\begin{equation}
    \mathcal{W}^l_Q = \nint[\Big]{\frac{\mathcal{W}^l}{S_w^l}} \in \texttt{INT}_\mathbf{n}, \qquad \mathcal{X}^l_Q = \nint[\Big]{\frac{\mathcal{X}^l}{S_x^l}} \in \texttt{INT}_\mathbf{n} \label{eq:quant}
\end{equation}
Where $S$ is the scaling factor determined by data precision $\mathbf{n}$ and the numerical range. Apparently, the integer-based data cannot be directly inserted back for training or optimization. In practice, the integer values will be scaled back to the floating-point domain as the ``dequantized~(DQ)'' values $\mathcal{W}^l_{\text{DQ}}$ and $\mathcal{X}^l_{\text{DQ}}$: 
\begin{equation}\label{eq:fake_q}
    \mathcal{W}^l_{\text{DQ}} = \mathcal{W}^l_Q\times S_w^l \in \texttt{Float}, \quad \mathcal{X}^l_{\text{DQ}} = \mathcal{X}^l_Q\times S_x^l \in \texttt{Float}
\end{equation}
In some cases, the integer-based zero point $\mathcal{Z}$ will also be part of the compression to shift the data range to formulate the signed or unsigned data, which is omitted in Eq.~\eqref{eq:fake_q} for simplicity.

Practically, the compressed weights and activation will be deployed to hardware with the low precision values only, while most quantization algorithms merge the quantize-and-dequantize altogether to ensure smooth gradient propagation. 
With the common convolution operation, we have:
\begin{align}
    &\text{Algorithm: }  \mathcal{Y}^l = \mathcal{W}^l_{\text{DQ}} \circledast \mathcal{X}^l_{\text{DQ}}\label{eq:alg_q} \\
    &\text{Hardware: }  \mathcal{Y}^l = (S_w^l S_x^l)\underbrace{\Big(\mathcal{W}^l_{\text{Q}} \circledast \mathcal{X}^l_{\text{Q}}\Big)}_{\text{Hardware Compatible}}\label{eq:hw_q}
\end{align}

As shown in Eq.~\eqref{eq:alg_q} and Eq.~\eqref{eq:hw_q}, the model pre-trained by the quantization algorithm cannot be deployed directly to hardware \JM{since the dequantized output is kept in high floating-point precision format}. 
Starting from the early exploration in \texttt{TensorFlow Lite}~\cite{jacob2018quantization} 
to the recent industry-level toolkit~\cite{siddegowda2022neural, gorbachev2019openvino}, prior works have attempted to automate the process from Eq.~\eqref{eq:alg_q} to Eq.~\eqref{eq:hw_q} with 8-bit precision, followed by parameter extraction or deployment to hardware. 

Meanwhile, the recent quantization algorithm mainly focuses on Eq.~\eqref{eq:alg_q}, where $\mathcal{W}^l_Q$ and $\mathcal{X}^l_Q$ can be compressed down to ultra-low precision~\citep{xiao2023smoothquant, zhang2022pokebnn}, power-of-two representation~\citep{Li2020Additive}, mixed precision with post-training quantization~(PTQ)~\citep{shang2023post, tu2023toward} or quantization-aware training~(QAT)~\citep{liu2022nonuniform, yang2019quantization}. 
Among all the recent algorithms, the dequantized $\mathcal{X}^l_{\text{DQ}}$ and $\mathcal{W}^l_{\text{DQ}}$ is essentially the ``discretized'' version of the original floating-point precision distribution. The hardware-readable $\mathcal{W}^l_{\text{Q}}$ and $\mathcal{X}^l_{\text{Q}}$ are treated as the intermediate results and \textbf{omitted} by the quantizer with in-place operation~\citep{li2021brecq,liu2022nonuniform}:
\vspace{-10pt}
\begin{figure}[h!]
    \centering
    \includegraphics[width=\linewidth]{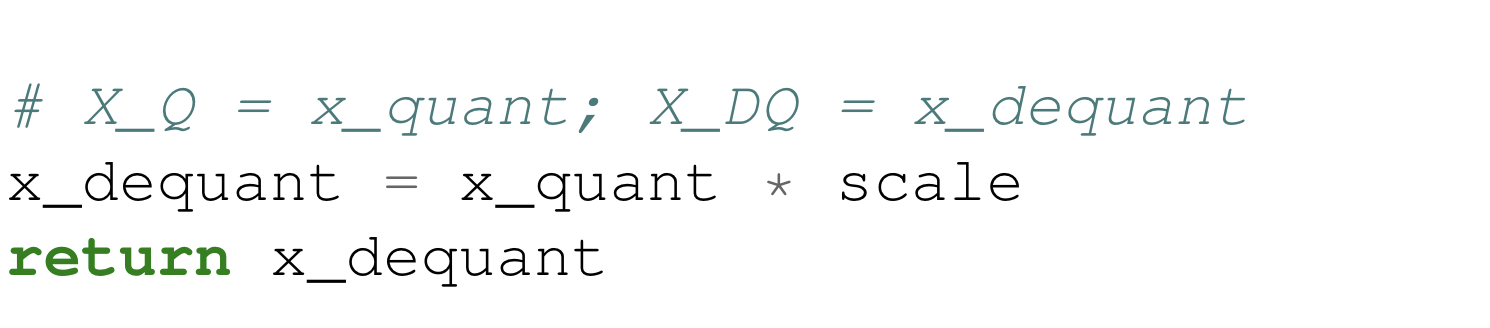}
\end{figure}
\vspace{-10pt}
With the model pre-trained by the quantization algorithms in PyTorch, \textbf{neither} $\mathcal{X}^l_{\text{Q}}$ or $\mathcal{W}^l_{\text{Q}}$ can be directly deployed for RTL-level verification. As a result, exploiting the benefits of the SoTA software-level quantization algorithm requires additional effort and conversion by the hardware designer. 

\subsection{Weight Sparsification}
Exploring sparsity for storage reduction and computation skipping has been widely investigated with various models, tasks, and granularities. 
In general, sparse training works can be categorized based on 1) the starting point of sparsification and 2) the granularity of sparsity. Specifically, post-training sparsification~\citep{dettmers2019sparse, evci2020rigging, liu2021sparse, yuan2021mest} removes the weights and then fine-tunes the model to recover the accuracy loss. Meanwhile, before-training~\cite{lee2018snip, wang2019picking} sparsification creates a sparse model prior to the main training procedure. 
Regarding the granularity, element-wise pruning~\citep{han2015deep} eliminates the individual weight value, while the recent works gradually expanded the spectrum, including hierarchical sparsity~\citep{kadetotad2020compressing, wu2023highlight}, hardware-specific structured sparsity~\citep{chu2020pim}, filter-wise sparsity~\citep{shen2022structural}, and structured fine-grained sparsity~\citep{zhou2021learning, zhang2022learning}. 
The pruning candidate under different granularities can be generated based on magnitude score~\citep{liu2021sparse, lee2018snip}, gradient score~\cite{liu2021sparse, yuan2021mest}, or evaluating the impact between adjacent layers or channels~\cite{Park2020Lookahead}. 

Despite the variety of granularity and pruning methods, the ultimate objective of sparsification is reducing weight storage and skipping the computation corresponding to the sparse weights and accelerating the on-device computation of hardware. Motivated by that, various accelerator designs have been proposed  on ASIC~\citep{Lee_JSSC19,Moon_ISSCC23,Conti_ISSCC23,Desoli_ISSCC23} or FPGA~\citep{meng2021fixyfpga, cai2023accelerating} implementation with various architectures. Although the current SoTA pruning algorithm can achieve up to 98\% element-wise sparsity~\citep{yuan2021mest, liu2021sparse}, most of the algorithms represent the sparsity by binary masks that are applied the \textbf{full-precision weights}. 
Given the persistent need to deploy low-precision models to prototype accelerators, isolating sparsification from quantization leads to limited practical benefits to hardware designers. 

\subsection{Self-supervised Pre-training}
In addition to the success of supervised deep learning, recent research works have demonstrated the strong visual representation learned via self-supervised learning~(SSL)~\cite{zbontar2021barlow, grill2020bootstrap, bardes2022vicregl}. In particular, the SSL-trained model outperforms the supervised learning counterpart in various downstream tasks. Starting from the early research with similarity-based sample alignment~\cite{chen2020simple} 
to the recent image reconstruction with masked input~\cite{he2022masked}, SSL is encouraging the model to learn the visual representation rather than the supervised label-matching. As a result, the pre-trained encoder achieves superior performance in the small-scale downstream vision tasks~(e.g., CIFAR, Flower), and outperforms the supervised learning counterpart. 

Compared to supervised training from scratch, the strong and versatile vision model pre-trained by SSL provides an alternative candidate for model compression and hardware deployment. Due to the elevated baseline performance, the SSL-trained model exhibits a stronger ``efficiency-accuracy'' tradeoff in the small-scale vision tasks. However, compressing the SSL-trained foundation vision model has not been considered as a candidate for compression or deployment in the mainstream deployment toolkits yet. 

\section{Torch2Chip}
\subsection{Hierarchical Design for Customized Quantization}
Different from the non-deployable computation path of most quantization algorithms, \texttt{Torch2Chip} employs the ``Dual-Path'' design, which separates the training and inference computation graph to support both floating-point-based ``fake quantization`` for training, and ``low precision-only'' computation for hardware verification and parameter extraction. Specifically, given the full precision weights \texttt{wfp} and the dequantized weights \texttt{wdq}, the training path of the \texttt{Base} quantizer is characterized as:
\begin{figure}[h!]
    \centering
    \includegraphics[width=\linewidth]{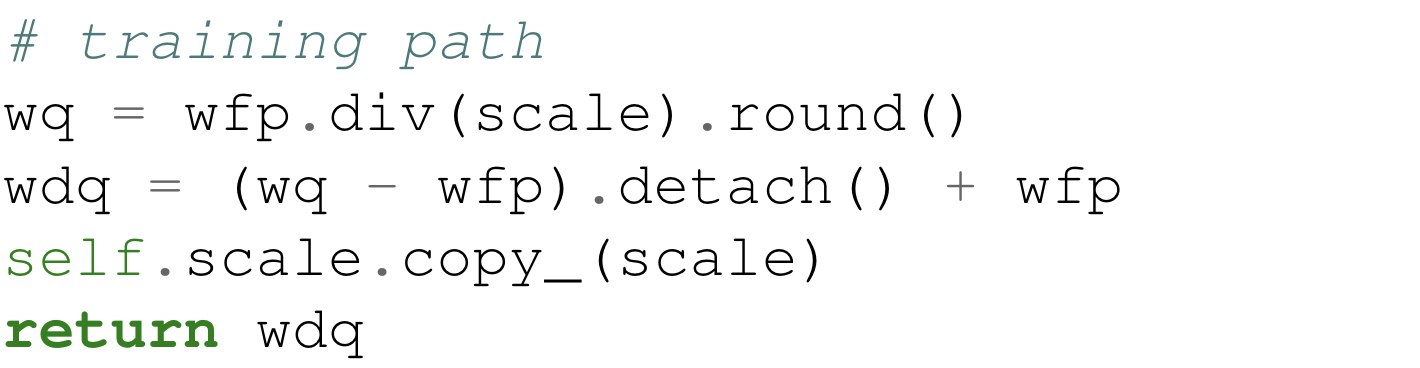}
\end{figure}
While the inference path focuses on low-precision parameters only which will be saved as the final state of the model: 
\vspace{-5pt}
\begin{figure}[h!]
    \centering
    \includegraphics[width=\linewidth]{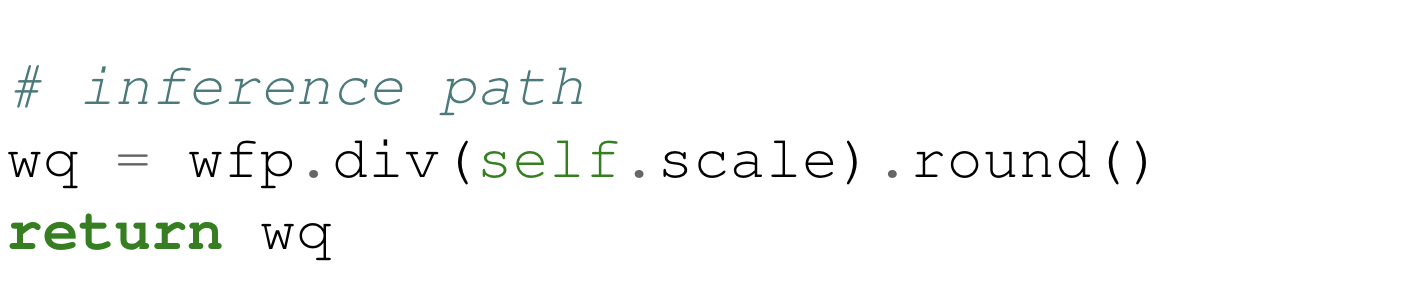}
\end{figure}
\vspace{-5pt}
The ``Dual-Path''-based computation scheme is designed as the \textbf{bottom-level logic} of the proposed \texttt{Torch2Chip} toolkit. On top of that, we define the \texttt{Base Module} and \texttt{Base Layer} as the starting point of the user-defined quantization method and computation, respectively. 
In particular, the customized quantization methods~(weight or activation) are constructed based on \texttt{\_QBase}, which holds the scaling factor and zero point as the registered parameters updated by the user-customized method in the training path. Subsequently, the customized quantizers are embedded into the second-level layer module to compress weights and activation. Following the same ``Dual-Path'' logic, the base layer modules also split the computation into inference and training paths separately to perform the same type of computation~(e.g., convolution) with different data types~(e.g., Integer or Float), as shown in Figure~\ref{fig:hierarchy}.

\begin{figure}[t!]
    \centering
    \includegraphics[width=0.6\linewidth]{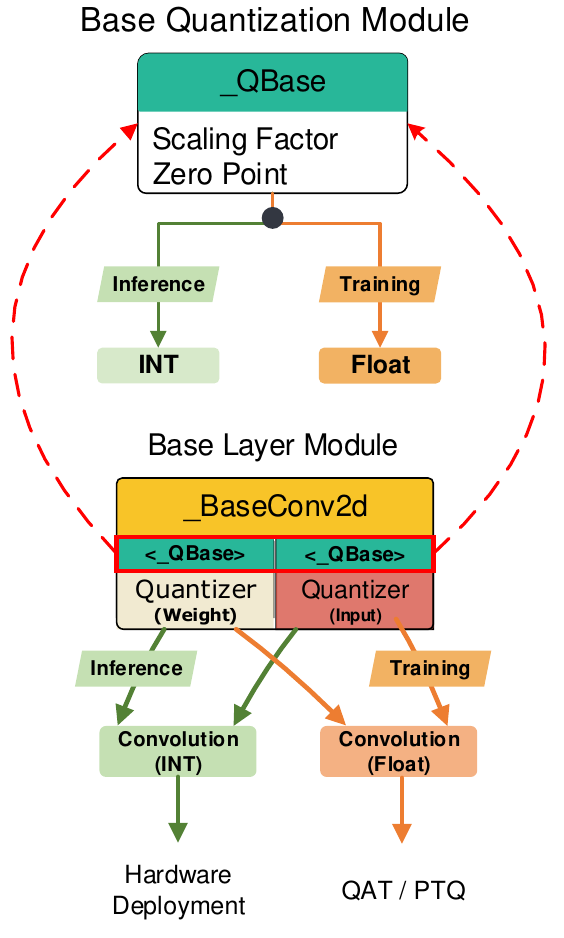}
        \vspace{-10pt}
    \caption{Hierarchical architecture of layer and module with the ``Dual-Path''-based design.}
    \label{fig:hierarchy}
\end{figure}

The proposed hierarchical workflow shows pragmatic benefits for algorithm customization and deployment. 
First, the independent training and inference paths can be switched globally on the top-level training or evaluation workflow. The separate computation graph ensures the training process will not be interfered by the integer-only non-differentiable computation. Secondly, the proposed hierarchical customization provided a generic and versatile template for customization. Since all the quantization parameters are registered inside the bottom-level module \texttt{\_QBase}, any kind of customized quantization can be converted correctly into low-precision only computation with the user-customized scaling factor and zero points. 

\begin{figure*}[t!]
    \centering
    \includegraphics[width=0.93\linewidth]{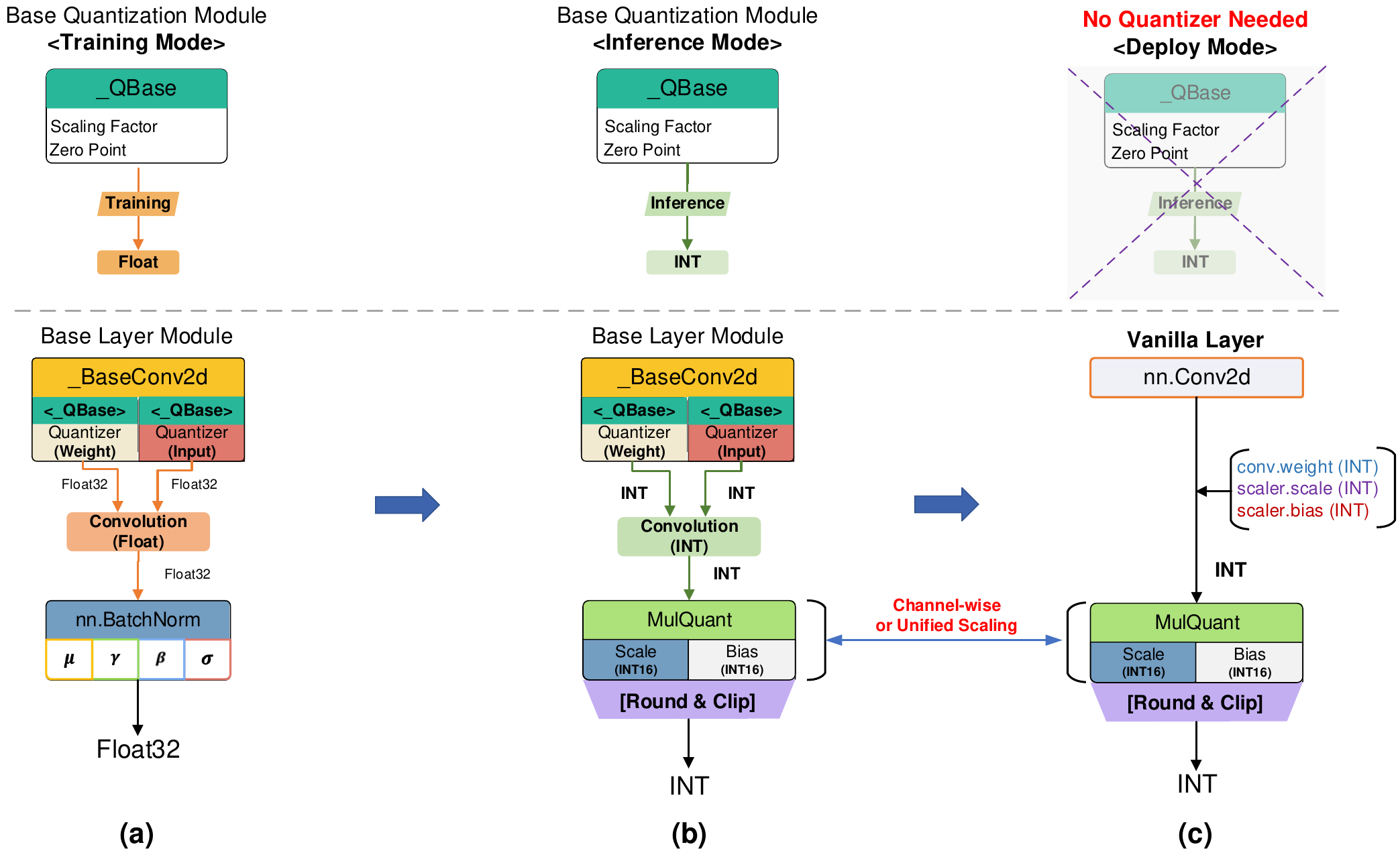}
        \vspace{-10pt}
    \caption{Training and fusion workflow of \texttt{Torch2Chip}: (a) training mode with fully customizable quantizer and layer, (b) inference mode with integer-only computation, and (c) deploy mode with integer-only parameters stored by the vanilla layer.}
    \label{fig:t2c_fuse}
    \vspace{-10pt}
\end{figure*}

\texttt{Torch2Chip} is also perfectly compatible with the recent \JM{adaptive quantization methods~\citep{yang2019quantization, nagel2020up, liu2023noisyquant}, while PyTorch~\citep{NEURIPS2019_9015} and OpenViNO~\citep{gorbachev2019openvino} mainly incorporate the
traditional nearest or stochastic rounding as the non-customizable option. For instance, }AdaRound~\citep{nagel2020up} learns the rounding threshold via a differentiable non-linear function $h$, where the integer weights are computed by adding the learnable offset $\alpha$: 
\begin{align}
    &\text{Training: } \mathcal{W}^l_Q = \nint[\Big]{\frac{\mathcal{W}^l}{S_w^l}} + h(\alpha) \\
    &\text{Inference: } \mathcal{W}^l_Q = \nint[\Big]{\frac{\mathcal{W}^l}{S_w^l}} + \{\mathbf{1} | \alpha \geq 0\}
\end{align}
To that end, the conventional scaling-based quantization in Eq.~\eqref{eq:quant} is not sufficient, and AdaRound~\citep{nagel2020up} is not directly compatible with PyTorch Quantization. Instead, \texttt{Torch2Chip} only focuses on the quantized parameter rather than the rounding process. The low-precision weights $\mathcal{W}^l_Q$ will be registered as an additional parameter inside the base layer for inference only, \JM{while the differentiable soft offset~($h(\alpha)$) will be added during the training path of \texttt{Torch2Chip}.} Although the training and inference are separately executed in \texttt{Torch2Chip}, users only need to implement the customized method in the training path, and the remaining steps will be executed \textbf{automatically} by \texttt{Torch2Chip}. 

\begin{figure*}[t!]
    \centering
    \includegraphics[width=0.93\linewidth]{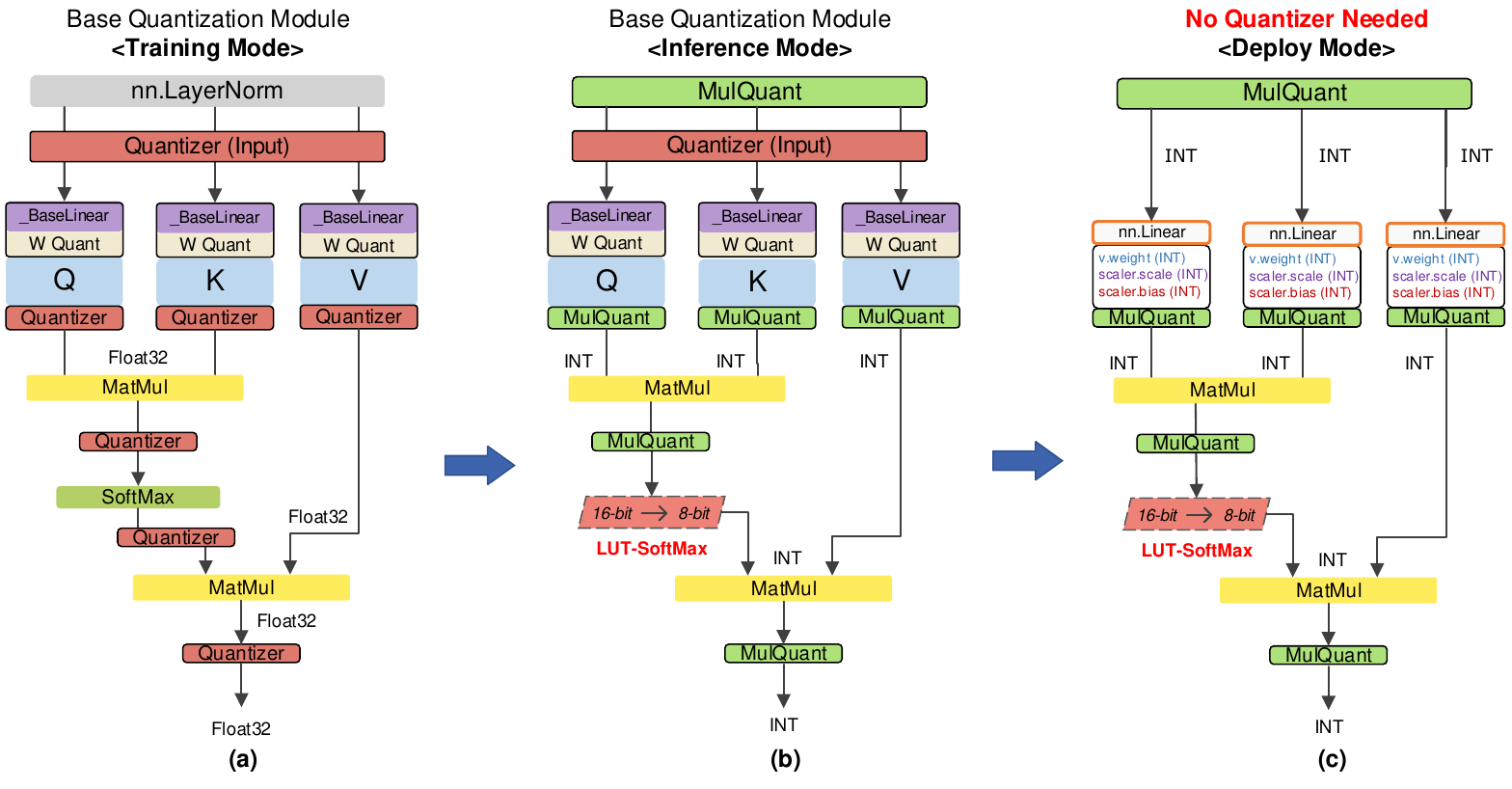}
        \vspace{-10pt}
    \caption{Training and fusion workflow of \texttt{Torch2Chip} with multi-head attention under 8-bit precision: (a) training mode with fully customizable quantizer and layer, (b) inference mode with integer-only computation and Look-up-Table~(LUT)-based Softmax function, and (c) deploy mode with integer-only parameters stored by the vanilla fully-connected layer.}
    \label{fig:t2c_vit}
\end{figure*}

\subsection{Fully Customizable Post-training Fusion}
\subsubsection{Fusing Quantizers and Normalization Layer}
As one of the most important components for deep neural network training, normalization layers (e.g., BatchNorm, LayerNorm) stabilize the training by scaling and shifting the distribution of the prior layer output, while separating the channel-wise difference with the weight and bias: 
\begin{equation}\label{eq:norm}
    \Bar{\mathcal{Y}^l} = \gamma \times \frac{\mathcal{Y}^l - \mu_s}{\sqrt{\sigma^2_s + \epsilon}} + \beta
\end{equation}
Where $\mu_s$ and $\sigma^2_s$ represent the mean and variance of $\mathcal{Y}^l$ along a certain dimension, and $\gamma$ and $\beta$ represent the learnable weight and bias of the normalization layer. 
The orientation of $\mu_s$ and $\sigma^2_s$ characterizes Eq.~\eqref{eq:norm} into different types of normalization methods~\citep{ioffe2015batch, ulyanov2016instance, ba2016layer, wu2018group}.

The common approach to compressing the normalization is fusing the parameter into weights prior to the convolution~(or matrix multiplication) and quantizing the resultant tensor altogether. Given the full precision weights $W^l$, such ``Pre-Fusing'' scheme can be represented as:
\vspace{-5pt}
\begin{align}
    &\mathcal{W}_{\text{Fuse}} = \frac{\gamma \mathcal{W}}{\sqrt{\sigma^2 + \epsilon}} \\
    &\mathcal{W}_Q = \nint[\Big]{\mathcal{W}_{\text{Fuse}} / S_w^l}
\end{align}
The normalization bias $\beta$ will be rearranged as the new bias term inside the layer~(convolution or linear):
\begin{align}
    &\Bar{\mathcal{Y}^l} = \mathcal{Y}^l + \beta^* \label{eq:pre_fuse} \\
    \text{where } \quad &\beta^* = \beta - \frac{\gamma \mu}{\sqrt{\sigma^2 + \epsilon}}
\end{align}
Although the ``Pre-Fusing''~\citep{jacob2018quantization, NEURIPS2019_9015} scheme has been adopted as the mainstream fusing strategy in the ML framework and toolkits, the normalization parameters are sensitive to low precision representation. In particular, the success of pre-fusion is built upon the 8-bit quantization, which has relatively high precision compared to the recent ultra-low precision algorithms~($<$8-bit). With the low-precision quantization, the ``Pre-Fusion'' scheme exhibits a high degree of instability and large degraded accuracy, as reported in both algorithm~\citep{park2020profit} and hardware design~\citep{meng2021fixyfpga}.

Alternatively, the normalization can be reformulated as the scaling and bias with respect to the channels of $\mathcal{Y}^l$:
\begin{align}
    &\Bar{\mathcal{Y}^l} = \gamma^*\times \mathcal{Y}^l + \beta^* \label{eq:post_scale} \\
    \text{where } \quad &\gamma^* = \frac{\gamma}{\sqrt{\sigma^2_s + \epsilon}}
\end{align}
Combining the fusion schemes in Eq.~\eqref{eq:pre_fuse} and Eq.~\eqref{eq:post_scale} with the quantization scaling in Eq.~\eqref{eq:hw_q}, we have:
\begin{align}
    &\text{8-bit:} \quad \mathcal{Y}^l_Q = \nint[\Big]{\frac{S_w^l S_x^l}{S_x^{l+1}}\Big(\mathcal{W}^l_{\text{Q}} \mathcal{X}^l_{\text{Q}}\Big) + \frac{\beta^*}{S_x^{l+1}}} \label{eq:sb_8bit} \\
    &\text{Sub 8-bit:} \quad \mathcal{Y}^l_Q = \nint[\Big]{\gamma^*\times\frac{S_w^l S_x^l}{S_x^{l+1}}\Big(\mathcal{W}^l_{\text{Q}} \mathcal{X}^l_{\text{Q}}\Big) + \frac{\beta^*}{S_x^{l+1}}} 
    \label{eq:sb_sub_8bit}
\end{align}
As shown in Eq.~\eqref{eq:sb_8bit} and Eq.~\eqref{eq:sb_sub_8bit}, both scenarios can be formulated as integer-only convolution followed by scaling and then adding bias. The major separation is the channel-wise scaling or unified scaling.
Different from the mainstream PyTorch and industry-standard toolkits \JM{that only support the unified scaling, Pre-Fusing, and 8-bit precision}, \texttt{Torch2Chip} incorporates \textbf{both} scaling scheme and supports the model fusion with respect to conventional 8-bit precision and the sub-8-bit quantization.

Unlike Pytorch~\citep{NEURIPS2019_9015} that keeps the unified scaling factor as high floating-point precision tensors, \texttt{Torch2Chip} automatically fuses the normalization module with the quantizer as the scaling and shifting into the high precision integer stored inside the \texttt{MulQuant} module with the user-defined integer and fractional precision. 
The ``Dual-Path'' computation graph and the automatic fusion logic ensure the user-defined quantization is deployable with integer-only parameters, as shown in 
Figure~\ref{fig:t2c_fuse} (a) (b).

\subsubsection{Integer-only Vision Transformer}
In addition to the automated compression and deployment on CNN,  \texttt{Torch2Chip} is also compatible with the compression on Vision Transformer~(ViT). Following the proposed ``Dual-Path'' computation graph and hierarchical customization, the multi-head attention and transformer block are compressed and fused automatically for integer-only inference and parameter extraction, as shown in Figure~\ref{fig:t2c_vit}(a) and Figure~\ref{fig:t2c_vit}(b). 

Different from the recent I-ViT~\citep{li2023vit}, which estimates the SoftMax function with shift-and-accumulation, \texttt{Torch2Chip} is equipped with Look-up-Table~(LUT)-based non-linear function approximation~(e.g., SoftMax, GeLU), which can be customized by users. The LUT-based approximation can enable an efficient hardware approximation. The non-linear function approximation and the accuracy impact have been largely ignored in the quantization of the mainstream DL frameworks~\citep{NEURIPS2019_9015, jacob2018quantization} with the na\"{i}ve full precision computation. 

\begin{figure*}[t!]
    \centering
    \includegraphics[width=\linewidth]{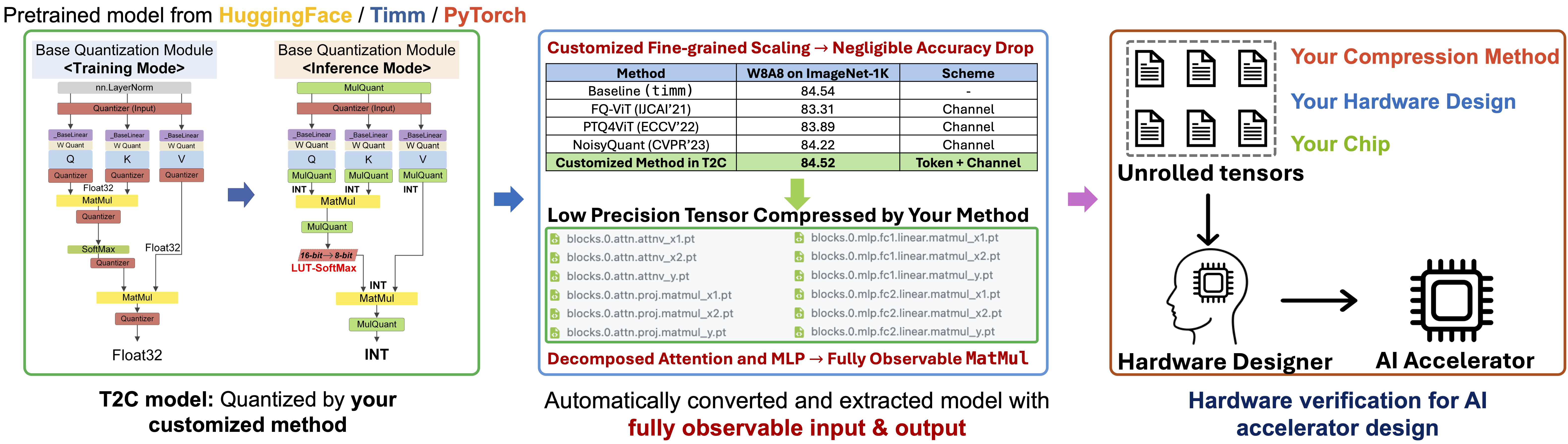}
    \caption{Automated and versatile parameter extraction with various output formats for RTL verification, integer-only model file, and native Pytorch quantization layer}
    \label{fig:t2c_export}
    \vspace{-10pt}
\end{figure*}

For the vision transformer with LayerNorm, $\mu$ and $\sigma$ are computed on the fly along the channel dimension. The serialized summation and averaging increase the latency compared to the pre-computed running statistics in BatchNorm. In \texttt{Torch2Chip}, LayerNorm is customizable with both instant statistics and pre-computed running statistics for different energy and latency requirements. 


\subsection{Compressing Powerful Foundation Vision Model}
Besides the versatile post-compression fusion and parameter extraction, \texttt{Torch2Chip} incorporates both supervised and self-supervised trainers for quantization~(QAT and PTQ) and powerful visual representation learning, respectively. 

\subsubsection{Self-supervised Pre-training}
In addition to the supervised learning compression, recent works have demonstrated the strong transfer learning performance of self-supervised learning~(SSL)~\citep{bardes2022vicregl, zbontar2021barlow, grill2020bootstrap}. With minimal fine-tuning, the SSL-trained backbone model outperforms the supervised learning~(fine-tuning) counterpart in various small-scale computer vision tasks. 
\texttt{Torch2Chip} incorporates the SSL in the pre-training phase and employs the powerful ``foundation model'' for subsequent compression and parameter extraction. 

In this work, we adopt the recent SoTA lightweight contrastive learning algorithm~\citep{meng2023slimmed} as the pre-training method. The model is trained by applying the cross-distillation~(XD) on top of the correlation-based contrastive learning loss~\cite{zbontar2021barlow}. Given the augmented input images $(A, \tilde{A})$, the cross-distillation is formulated as:
\begin{align}\label{eq:cd_loss}
    &\mathcal{L}_{\texttt{XD}}^A = \sum_i(1-C_{ii}^{A\tilde{A}}) + \lambda \sum_i\sum_{i\neq j}(C_{ij}^{A\tilde{A}})^2
\end{align}
Where $C_{ij}$ represents the correlation between the different dimensions of the latent output $(z, \tilde{z})$ encoded from $(A, \tilde{A})$ with the neural network model. Together with the Barlow Loss~\citep{zbontar2021barlow}, the DNN model is trained to learn a robust visual representation from the input, which further leads to strong transfer learning performance for various small-scale vision tasks. 
Unlike other model compression toolkits~\cite{gorbachev2019openvino, siddegowda2022neural} that employ supervised training only, \JM{\texttt{Torch2Chip} incorporates self-supervised learning as an alternative and high-performance pre-training option to users. The pre-trained model can be further fine-tuned and compressed in the downstream tasks to 
facilitate the performance. All the training and fine-tuning schemes are customizable in the \texttt{TRAINER} selection that is introduced in Section~\ref{sec:auto}.}

\begin{table*}[t!]
\centering
\caption{ImageNet-1K accuracy comparison between \texttt{Torch2Chip} and the recent DNN deployment toolkits.}
\vspace{3pt}
\resizebox{0.83\linewidth}{!}{\begin{tabular}{cccccccc}
\toprule
\textbf{Toolkit}                                                                             & \textbf{Method}                                                                 & \textbf{Training Method} & \textbf{Model}     & \textbf{W/A (bit)}  & \textbf{Scale and Bias (INT, Frac)} & \textbf{Accuracy (\%)} & \textbf{Customizable} \\ \midrule
\begin{tabular}[c]{@{}c@{}}AIMET (Qualcomm) \\ \citep{siddegowda2022neural}\end{tabular}                                                                               & \begin{tabular}[c]{@{}c@{}}AdaRound\\ \citep{nagel2020up}\end{tabular}                                                               & PTQ             & ResNet-50 & 8/8 & Float                      & 75.45 (-0.55)         &  \xmark            \\ \midrule
\begin{tabular}[c]{@{}c@{}}OpenVINO (Intel) \\ \citep{gorbachev2019openvino}\end{tabular}                                                                            &  MinMax Quant.                                                                       & PTQ             & ResNet-50 & 8/8 & Float                      & 75.98~(0.02)              &  \xmark            \\ \midrule
\rowcolor{ora}\textbf{Torch2Chip (Ours)}                                                                   & \begin{tabular}[c]{@{}c@{}}QDROP\\ \citep{wei2022qdrop}\end{tabular}                                                                  & PTQ             & ResNet-50 & \textbf{4/4} & \textbf{INT (12, 4)}                & 74.40 (-1.60)         &   \cmark           \\ \midrule
\rowcolor{ora}\textbf{Torch2Chip (Ours)}                                                                   & \begin{tabular}[c]{@{}c@{}}QDROP\\ \citep{wei2022qdrop}\end{tabular}                                                                  & PTQ             & ResNet-50 & 8/8 & \textbf{INT (12, 4)}                & \textbf{75.96 (-0.04)}         &    \cmark          \\ \bottomrule
\end{tabular}}
\label{tbl:compare}
\vspace{-10pt}
\end{table*}

\begin{table*}[t!]
\centering
\caption{CIFAR-10 accuracy comparison with the integer-only DNN model compressed by \texttt{Torch2Chip} with customized quantization.}
\vspace{3pt}
\resizebox{0.85\linewidth}{!}{\begin{tabular}{cccccccc}
\toprule
\textbf{Method}                                                                                          & \textbf{Model}        & \textbf{Training Method} & \textbf{W/A (bit)}  & \textbf{\# of Param. (M)} & \textbf{Scale and Bias (INT, Frac)} & \textbf{Accuracy (\%)} & \textbf{Model Size (MB)} \\ \midrule
\begin{tabular}[c]{@{}c@{}}SAWB + PACT\\ \citep{choi2019accurate}\end{tabular} & ResNet-20    & QAT             & \textbf{2/2} & 0.27             & INT (13, 3)                & 90.22 (-1.17) & 0.07            \\ \midrule
\begin{tabular}[c]{@{}c@{}}SAWB + PACT\\ \citep{choi2019accurate}\end{tabular}                 & ResNet-20    & QAT             & \textbf{4/4} & 0.27             & INT (13, 3)                & 91.24 (-0.73) & 0.14            \\ \midrule
\begin{tabular}[c]{@{}c@{}}RCF \\ \citep{Li2020Additive}\end{tabular}                          & ResNet-18    & QAT             & \textbf{4/4} & 11.17            & INT (12, 4)                & 94.56 (-0.21) & 5.59            \\ \midrule
\begin{tabular}[c]{@{}c@{}}RCF\\ \citep{Li2020Additive}\end{tabular}                           & ResNet-18    & QAT             & 8/8 & 11.17            & INT (12, 4)                & 94.77 (-0.01) & 11.17           \\ \midrule
\begin{tabular}[c]{@{}c@{}}RCF\\ \citep{Li2020Additive}\end{tabular}                           & ViT-7        & QAT             & 8/8 & 6.3              & INT (13, 3)                & 89.63 (-0.02) & 6.30            \\ \midrule
\begin{tabular}[c]{@{}c@{}}PROFIT\\ \citep{park2020profit}\end{tabular}                        & MobileNet-V1 & QAT             & \textbf{4/4} & 4.2              & INT (12, 4)                & 89.42 (-0.35) & 2.10            \\ \midrule
\begin{tabular}[c]{@{}c@{}}PROFIT\\ \citep{park2020profit}\end{tabular}                        & MobileNet-V1 & QAT             & 8/8 & 4.2              & INT (12, 4)                & 89.73 (-0.01) & 4.20            \\ \midrule
\begin{tabular}[c]{@{}c@{}}AdaRound\\ \citep{nagel2020up}\end{tabular}                         & MobileNet-V1 & PTQ             & 8/8 & 4.2              & INT (12, 4)                & 89.57 (-0.17) & 2.10            \\ \midrule
\begin{tabular}[c]{@{}c@{}}PyTorch Quant.\\ \citep{NEURIPS2019_9015}\end{tabular}             & MobileNet-V1 & PTQ             & 8/8 & 4.2              & Float32                    & 89.34 (-0.40) & 4.20            \\ \bottomrule
\end{tabular}}
\label{tbl:cifar_res}
\vspace{-10pt}
\end{table*}




\subsection{Automated and Customized Parameter Extraction}\label{sec:auto}
On top of all the automatic layer-wise fusion, the final phase of \texttt{Torch2Chip} is converting the user-customized low-precision model into the integer-only representation with the native vanilla layers of PyTorch~\citep{NEURIPS2019_9015}.

Since all the operations are converted to integer-based operations after the automated fusion, \texttt{Torch2Chip} removes the user-customized quantizer and assigns the low-precision integer weights to the vanilla layers of Pytorch.
The scaling and normalization are fused into the \texttt{MulQuant}, as shown in Figure~\ref{fig:t2c_fuse}(c) and Figure~\ref{fig:t2c_vit}(c). By doing so, the vanilla models are customized and compressed into the low-precision model while keeping the same architecture as the original model. 
Instead of saving the full-precision weight parameters and executing the quantization on the fly, our ``\textit{vanilla-custom-vanilla}'' scheme leads to multiple benefits for the subsequent hardware deployment:
\vspace{-10pt}
\begin{itemize}
    \item Keeping the vanilla model architecture as the output of \texttt{Torch2Chip} ensures the compressed parameters are saved in the same file format as the original model. As a result, \texttt{Torch2Chip} guarantees the ``real compression'' with generic modules of PyTorch.
    \item The integer-only model file is compatible with various data formats, including both decimal and hexadecimal representation. 
    From the perspective of prototyping, the customized compression methods are deployable to hardware in a simple and automated manner.
    \item Different from other open-sourced quantization algorithms, the vanilla layer and the \texttt{MulQuant} layers can be packed into the native quantization layer of PyTorch~\cite{NEURIPS2019_9015}, which will be saved as the default \texttt{torch.qint} format.  
\end{itemize}

At the top level, \texttt{Torch2Chip} automates the entire compression workflow with a simple and elegant execution process consisting of \textbf{five lines of code only}: 
\vspace{-5pt}
\begin{figure}[h!]
    \centering
    \includegraphics[width=\linewidth]{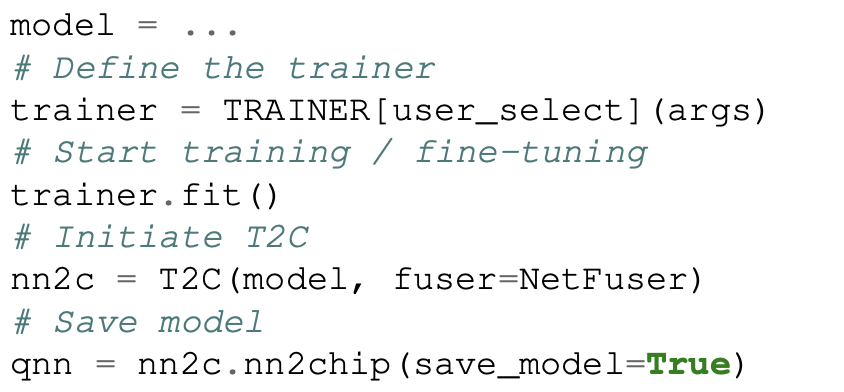}
\end{figure}
\vspace{-5pt}

\JM{\texttt{Torch2Chip} covers a wide spectrum of training schemes for different compression algorithms (\texttt{TRAINER}), including supervised QAT, PTQ, sparse training, and self-supervised foundation vision model training. The detailed settings will be elaborated in the official document of \texttt{Torch2Chip}. }

Overall, \texttt{Torch2Chip} successfully merges the gap between the customized SoTA algorithm, DL framework, and RTL-based chip prototyping while minimizing the additional programming effort of hardware designers. Figure~\ref{fig:t2c_export} shows the high versatility of \texttt{Torch2Chip} with various export formats.

\section{Experimental Results}\label{sec:exp_res}
This section reports the comprehensive performance evaluation of the proposed \texttt{Torch2Chip}. We evaluate the proposed toolkit with ResNet~\citep{he2016deep}, MobileNet~\citep{howard2017mobilenets}, and Vision Transformer~(ViT)~\citep{dosovitskiy2020image} against various compression 
methods. We employ multiple recent quantization and sparsification algorithms based on the proposed workflow to demonstrate the high degree of customization.

\begin{table*}[t!]
\centering
\caption{Sparse and low precision ResNet-50 compressed by \texttt{Torch2Chip} with customized methods on ImageNet-1K dataset.}
\vspace{3pt}
\resizebox{0.9\linewidth}{!}{\begin{tabular}{ccccccc}
\toprule
\textbf{Method}    & \textbf{W/A (bit)} & \textbf{Quantization} & \textbf{Model}     & \textbf{W/A (bit)} & \textbf{Sparsity (\%)} & \textbf{Accuracy (\%)}   \\ \midrule
\begin{tabular}[c]{@{}c@{}}GraNet \citep{liu2021sparse}\end{tabular}     & 8/8       & PTQ          & ResNet-50 & 8/8 & 80\%          & 75.15 (-0.85\%) \\ \midrule
\begin{tabular}[c]{@{}c@{}}GraNet \citep{liu2021sparse}\end{tabular}    & 4/4       & PTQ          & ResNet-50 & 4/4 & 80\%          & 73.38 (-2.62\%) \\ \midrule
N:M = 2:4~\citep{zhou2021learning} & 8/8       & PTQ          & ResNet-50 & 8/8 & 50\%          & 75.44 (-0.75\%) \\ \midrule
N:M = 2:4~\citep{zhou2021learning} & 4/4       & PTQ          & ResNet-50 & 4/4 & 50\%          & 74.16 (-1.84\%) \\ \bottomrule
\end{tabular}}
\label{tbl:t2c_sparse}
\end{table*}

\begin{table*}[t!]
\centering
\caption{Transfer fine-tuning of MobileNet~\citep{howard2017mobilenets} pretrained by the proposed method.}
\vspace{3pt}
\resizebox{0.9\linewidth}{!}{\begin{tabular}{cccccccccc}
\toprule
\textbf{Method}    & \textbf{Encoder}   & \textbf{W/A (bit)}            & \textbf{CIFAR-10} & \textbf{CIFAR-100} & \textbf{Aircraft} & \textbf{Flowers} & \textbf{Food-101}  \\ \midrule
Torch2Chip + Supervised Learning + PTQ & Mob-V1 (1$\times$)    & 8/8         &    89.74               & 65.98                   &  60.09                 &  72.23                & 56.41                  \\ \midrule
\rowcolor{ora}\begin{tabular}[c]{@{}c@{}}\textbf{Torch2Chip + XD + PTQ}\\ \citep{meng2023slimmed}\end{tabular}           & Mob-V1 (1$\times$)  & 8/8            & \textbf{94.37}                  & \textbf{74.29}                   & \textbf{68.44}             &  \textbf{86.42}                &  \textbf{70.21}            \\ \bottomrule
\end{tabular}}
\label{tbl:downsteram}
\end{table*}

\subsection{Integer-only DNN on ImageNet-1K Dataset}
\texttt{Torch2Chip} is also compatible with the large-scale ImageNet dataset. In addition to the re-implementation of the SoTA quantization algorithms based on \texttt{Torch2Chip}, we evaluate the versatility of \texttt{Torch2Chip} by comparing it with the native PyTorch quantization API~\citep{NEURIPS2019_9015} and the recent industry-standard toolkits~\citep{siddegowda2022neural, gorbachev2019openvino}. 
Table~\ref{tbl:compare} summarizes the PTQ performance on ResNet-18 compressed by \texttt{Torch2Chip} on the ImageNet-1K dataset with 8-bit and 4-bit model precision quantized by QDrop~\cite{wei2022qdrop} under the workflow of \texttt{Torch2Chip}. The fully customizable and observable compression workflow enables the freedom of employing high-performance algorithms with sub-8-bit precision while keeping the integer-only computation. The starting point of the PTQ is the full-precision ResNet-50 model pre-trained by PyTorch. The fused scaling factor and bias are quantized to INT16 fixed-point representation with 12 fractional bits and 4 integer bits.

Compared to other recent deployment toolkits~\citep{siddegowda2022neural, gorbachev2019openvino}, the objective of \texttt{Torc2Chip} is ``customize a compressed model'' rather than ``getting a compressed model''. The full stack of customization and the automated model conversion enables the hardware designers to implement the algorithm freely and deploy the model directly. 

\subsection{Integer-only DNN on CIFAR-10 Dataset}
We also evaluate the compressed DNN model performance on the CIFAR-10 dataset and summarize the results in Table~\ref{tbl:cifar_res}. 
For the QAT-based algorithms, the CNN models are trained by 200 epochs from scratch with SGD optimizer. The learning rate schedule and the required regularization are reimplemented based on the parameter settings reported in the original paper. Table~\ref{tbl:cifar_res} reports the optimal scaling precision with the best accuracy. 

Besides the fully customizable compression scheme for ultra-low precision, \texttt{Torch2Chip} performs better than PyTorch quantization with pre-defined 8-bit precision. 
With the MobileNet-V1 model, \texttt{Torch2Chip} outperforms the PyTorch native quantization scheme without introducing any floating point operations, as presented in Section~\ref{sec:exp_res}.

\subsection{Sparse Training with User-defined Sparsity}
Besides the quantization, \texttt{Torch2Chip} also allows users to exploit sparsity with the user-customized sparsification method. Starting from the basic pruner with element-wise sparsity~\citep{han2015deep}, \texttt{Torch2Chip} provides the customized example pruner with N: M sparsity~\citep{zhou2021learning}. Table~\ref{tbl:t2c_sparse} summarizes the performance of the sparse ResNet-50 pruned by Torch2Chip with different granularity. Starting from scratch, the dense model is pruned with gradually increased sparsity, and the post-training quantization is applied to compress the full-precision sparse model into integer-only representation. 
Instead of simply representing the sparsity as an additional mask, the sparsified and converted models save the pruned parameter as the raw zero values in the resultant integer model, which can be saved and exported directly to the user-defined file format. 

\subsection{Compressed Transfer Learning based on Self-supervised Pre-training}
Different from~\citep{siddegowda2022neural, gorbachev2019openvino} with supervised pre-training only, \texttt{Torch2Chip} incorporates the powerful SSL-trainer with cross-distillation (XD)~\citep{meng2023slimmed} to achieve the superior performance on the downstream small-scale vision tasks. 
Before quantizing the model, the full-precision MobileNet-V1 is pre-trained by the self-supervised \texttt{Trainer} on ImageNet-1K. The pre-trained foundation model is fine-tuned and compressed with the supervised \texttt{Trainer} on various downstream tasks. 

Table~\ref{tbl:downsteram} shows a transfer learning example of SSL-trained MobileNet-V1 compressed by \texttt{Torch2Chip} with 8-bit precision.
Compared to supervised learning from scratch, the SSL-trained MobileNet-V1 achieves \textbf{94.37\%} and \textbf{78.22\%} CIFAR-10 and CIFAR-100 accuracy, which largely outperforms the supervised learning baseline with 4.63\% and 12.2\% accuracy improvements. \JM{The entire workflow of training, fine-tuning, and compression is fully customizable by users with different datasets and precision.} The resultant high-performance models are automatically converted into integer-only computation, which can be exported to various formats as shown in Figure~\ref{fig:t2c_export}.

\section{Conclusion}
In this paper, we presented \texttt{Torch2Chip}, a fully customizable, fully observable development toolkit designed for prototype DNN hardware accelerator design. Unlike other works with limited customization or non-extractable parameters, our work enables end-to-end customization while automating post-compression fusion and parameter extraction with different output formats. \texttt{Torch2Chip} bridges the research gaps between algorithm, hardware designer, and DL frameworks and further enriches the training method with powerful self-supervised learning. The automated deployment process accelerates the design process of prototyping with ASIC or FPGA platforms. The open-sourced toolkit will be updated continuously to support full-stack customization with different deep-learning model deployment scenarios. 

\section{Acknowledgement}
This work was supported in part by Samsung Electronics and the Center for the Co-Design of Cognitive Systems (CoCoSys) in JUMP 2.0, a Semiconductor Research Corporation (SRC) Program sponsored by the Defense Advanced Research Projects Agency (DARPA).

\newpage
\bibliography{t2c} 
\bibliographystyle{mlsys2024}



\end{document}